\documentclass[aps,preprint,preprintnumbers,nofootinbib,showpacs,prd]{revtex4}
\usepackage{graphicx,color,amsmath,amsfonts}

\newcommand{\beq}     {\begin{equation}}
\newcommand{\eeq}     {\end{equation}}
\newcommand{\bea}     {\begin{eqnarray}}
\newcommand{\eea}     {\end{eqnarray}}
\newcommand{\bad}{\begin{array}{ccc}}
\newcommand{\ea}{\end{array}}
\newcommand{\me}{\,\rlap{/}{\!E}}
\newcommand{\met}{\,\rlap{/}{\!E_T}}

\newcommand{\lsim}{\mathrel{\mathop{\kern 0pt \rlap
  {\raise.2ex\hbox{$<$}}}
  \lower.9ex\hbox{\kern-.190em $\sim$}}}
\newcommand{\gsim}{\mathrel{\mathop{\kern 0pt \rlap
  {\raise.2ex\hbox{$>$}}}
  \lower.9ex\hbox{\kern-.190em $\sim$}}}
  
\newcommand{\no}     {\nonumber}

\newcommand{\br}{{\rm Br}}
\newcommand{\neut}{{\widetilde{\chi}^0_2}}
\newcommand{\cha}{{\widetilde{\chi}^\pm_1}}
\newcommand{\chap}{{\widetilde{\chi}^+_1}}
\newcommand{\cham}{{\widetilde{\chi}^-_1}}
\newcommand{\bino}{{\widetilde{B}}}

\newcommand{\gm}      {\gamma}
\newcommand{\Gm}      {\Gamma}

\newcommand{\ppb}      {p \bar{p}}
\newcommand{\fb}{{\,{\rm fb}}}
\newcommand{\fbi}{{\,{\rm fb}^{-1}}}
\newcommand{\pbi}{{\,{\rm pb}^{-1}}}

\newcommand{\mev}{{\;{\rm MeV}}}
\newcommand{\gev}{{\;{\rm GeV}}}
\newcommand{\tev}{{\;{\rm TeV}}}

\newcommand{\ax}      { {\tilde{a}} }
\newcommand{\grno}      { {\widetilde{G}} }
\newcommand{\gno}      { {\tilde{g}} }

\begin{document}
\title{
Probing axino LSP \\
from diphoton events with large missing transverse
energy
}

\author{Sanghyeon Chang}
\email{sang.chang@gmail.com}

\author{Kang Young Lee}
\email{kylee14214@gmail.com}

\author{Jeonghyeon Song}
\email{jhsong@konkuk.ac.kr}

\affiliation{
Division of Quantum Phases \& Devices, School of Physics, 
Konkuk University, Seoul 143-701, Korea
}

\date{\today}
\begin{abstract}
In a supersymmetry model
with an axino as the
lightest supersymmetric particle (LSP) 
and a Bino as the next LSP (NLSP),
supersymmetric particle production
ends up with including two Binos, followed by each Bino's 
decaying into a photon and an axino.
Final states are diphoton with large missing energy.
We have comprehensively studied
the implication
of $\gm\gm+\me_{(T)}$ data from the ALEPH, CDF II,
ATLAS and CMS experiments.
No excess over the standard model backgrounds
can be explained in this model if the Bino NLSP 
decays outside the detector.
Long life time of the Bino  is possible because 
of high Peccei-Quinn symmetry breaking scale $f_a$.
The ALEPH and CDF II data put a very strong bound on
$f_a$ for light Bino case with $m_\bino \lsim 150\gev$:
the narrow hadronic axion window around $f_a \sim 10^6$ GeV
is completely closed.
The recent ATLAS and CMS data show very interesting exclusion
$f_a \gsim 10^5$ GeV
for the Bino mass below 700 GeV.
This is already stronger than the previous laboratory bounds.
\end{abstract}
\maketitle

%%%%%  TEXT BEGINS  %%%%%%%

Impressive performance of the LHC in its early operation
escalates our expectation to reveal the secrets of the universe.
One of the most profound mysteries is 
the identity of the dark matter
(DM).
In particle physics, the DM is usually explained by 
a weakly interacting stable particle 
with a well-motivated symmetry 
for new physics beyond the standard model (SM).
If this DM particle is produced at a high energy collider,
it escapes detection and leaves the signal of
missing energy.
In order to obtain the missing energy information,
we need to measure the four-momenta of accompanying SM particles.
An isolated photon is a good candidate for this role. 

Recently the ATLAS and CMS collaborations have reported the
search for diphoton events 
with large missing transverse 
energy  $\met$
based on the LHC data 
at $\sqrt{s}=7\tev$~\cite{Aad:2011kz,arXiv:1111.4116,arXiv:1103.0953,EXO-11-067}.
No excess above the SM 
backgrounds has been found.
Together with the previous results from 
 LEP2~\cite{Heister:2002ut,Heister:2002vh,Abdallah:2003np} and
CDF II~\cite{Aaltonen:2008dm,Aaltonen:2009tp}, 
these events constrain
new physics models which predict $\gm\gm+\met$ signal.
One good example is the minimal supersymmetric 
standard model (MSSM)
with the gauge mediated supersymmetry breaking (GMSB).
In this model, the light gravitino $\grno$ is the 
lightest supersymmetric particle (LSP).
A supersymmetric (SUSY) particle eventually decays
into the next-LSP (NLSP), and 
the NLSP sequentially decays into
a gravitino and a SM particle. 
In most parameter space,
the NLSP is the Bino,
and the Bino decays almost exclusively into a photon 
and a gravitino.
With $R$ parity conservation, 
SUSY particles are always produced in pairs.
Therefore all the SUSY final states include two photons 
plus missing energy carried by two gravitinos. 
In this regard, the experiments of Ref.~\cite{Aad:2011kz,
arXiv:1111.4116,arXiv:1103.0953,EXO-11-067,Aaltonen:2008dm,
Aaltonen:2009tp,Heister:2002ut,Heister:2002vh,Abdallah:2003np}  
provide bounds on the lifetime and mass of the Bino
in the GMSB models~\cite{Allanach:2002nj,Meade:2008wd,Buican:2008ws}. 
%But the bound is approximately valid to any MSSM like model with Bino as a NLSP and gravitino as a LSP.

Another interesting new physics model for the diphoton events
plus large $\met$ is the SUSY model 
with axino LSP~\cite{axino}. 
This model is motivated in order to solve the strong CP problem.
It introduces 
an axion superfield with 
the Peccei-Quinn (PQ) symmetry.
In addition to the MSSM particle contents,
we have an axion and its superpartner axino.
This axino can be very light with mass about $10\mev$~\cite{no-scale1,no-scale2}.
Then as in the GMSB model, the NLSP Bino
decays into an axino and a photon.
The final states include two photons with missing energy.
The $\gm\gm+\met$ data 
can have significant implications on this axino-LSP and Bino-NLSP model.
This is our main goal.

Let us begin with a brief review on the axion field.
The strong CP problem arises 
from the strong CP odd term of
\begin{align}
{\cal L}_\theta= \frac{\theta}{32\pi^2} 
G^a_{\mu\nu}\widetilde G^a_{\mu\nu}
,
\end{align}
where $G^a_{\mu\nu}$ is the field strength of a gluon.
The absence of neutron electric dipole moment 
leads to extremely small value of $\theta$:
$|\theta|\lsim 0.7 \times 10^{-11}$ \cite{arXiv:0807.3125}.
It requires to be explained by some symmetry argument.
One way is replacing $\theta$ as a dynamical field $\theta(x)=a(x)/f_a$, 
where $a(x)$ is an extremely weakly interacting pseudo-scalar field, 
called an axion, and $f_a$ is the axion decay constant. 
The effective Lagrangian for the axion field is
\begin{align}
{\cal L}_a= \frac{1}{2}(\partial_\mu a)^2
    +\frac{a}{f_a}\left(\frac{g_s^2}{32\pi^2} 
G^{a\mu\nu}\widetilde G^a_{\mu\nu}
    +C_{a\gamma}\frac{e^2}{32\pi^2} 
F^{\mu\nu}\widetilde F^{\mu\nu}\right) ,
\end{align}
where 
$|C_{a\gamma}|\sim 1$ is a model-dependent parameter.
%, and
%$aW\tilde W$ and $aZ\tilde Z$ couplings are orthogonal 
%to axion-photon couplings.

After the QCD phase transition, the axion acquires the mass of
\begin{equation}
 m_a =\frac{\sqrt{z}}{1+z}\frac{f_\pi m_\pi}{f_a},
\end{equation}
where $z=m_u/m_d$, $f_\pi$ is the pion decay constant,
and $m_\pi$ is the pion mass. 
The axion-photon interaction can be rewritten as
\bea
L_{a\gamma\gamma}=\frac{g_{a\gamma\gamma}}{4}a
F_{\mu\nu}{\widetilde F}^{\mu\nu},
\eea
where 
$g_{a\gamma\gamma}=\alpha C_{a\gamma}/(2\pi f_a)$.

In the literature,
there are two popular models for a very light and weakly interacting axion,
the hadronic axion model~\cite{Kim:1979if,Shifman:1979if} 
and the DFSZ axion 
model~\cite{Zhitnitsky:1980tq,Dine:1981rt}.
Current laboratory bound on the axion decay constant is 
$f_a \gsim 10^4 \gev$.
Much stronger bound comes from
the astrophysical and cosmological searches such that
$ 10^9 \gev \lsim f_a \lsim 10^{12}\gev$~\cite{arXiv:0807.3125,Freitas:2011fx}.
Below this cosmological bound,
still survives a
very narrow but interesting range 
around $f_a \sim 10^6 \gev$, called the hadronic axion window~\cite{Chang:1993gm}.

%The bound from bino decay into axino and photon is not strong 
%compared with astrophysical and cosmological bound. 
%However it can reach the hadronic axion window.

We consider the hadronic axion model with supersymmetry~\cite{Hisano:1996ja}.
In our model, the axion interaction is described 
by the following superpotential:
\begin{align}
 W=y\, \Phi  Q_1  Q_2,
\end{align}
where $\Phi=\phi+\sqrt{2}\chi \theta+F_\Phi \theta\theta$ 
is the axion superfield, and 
$Q_{1,2}$ are $SU(2)_L$ singlet heavy quark superfields.
The vacuum expectation value $\langle \phi \rangle=f_a/\sqrt{2}$ 
is the PQ symmetry breaking scale.
The complex field $\phi$ consists of
the real-scalar saxion field $s$ and the pseudo-scalar
axion field $a$.
The axino field is
\begin{equation} 
\tilde{a}=\left(
\begin{array}{l}\chi \\ \bar\chi 
\end{array}
\right).
\end{equation}
 
The mass of axino is model-dependent. 
An interesting possibility connected with the DM
is that the axino can be
significantly lighter than other SUSY particles,
and becomes the LSP~\cite{Rajagopal:1990yx,Chun:1992zk}.
In the no-scale supergravity model, for example,
the axino mass is generated through one-loop diagrams:
\bea
 m_{\tilde a}\simeq \frac{1}{16\pi^2}\,y^2 \,m_{\rm SUSY},
\eea
where $m_{\rm SUSY}$ is the induced SUSY breaking soft mass.
If $y\simeq 0.1$ and $m_{\rm SUSY} \simeq 100\gev$,
we have $m_{\tilde a} \sim 10 \mev$.
If the axino is the LSP, 
then the NLSP will decay into an axino and a SM particle.
Hereafter we consider the case where 
the axino is the LSP and the Bino is the NLSP.

The axino field has the interaction vertices with $\tilde{g} g$ and $\bino \gm$ at one-loop level,
mediated by  two $SU(2)_L$ singlet heavy quarks $U$ and $D$
in a simple model.
The  electromagnetic charges of $U$ and $D$
are  $2/3$ and $-1/3$
respectively.
The effective Lagrangian is 
\begin{align}
 {\cal L}_{\tilde a\tilde g g}
       &=\frac{\alpha_s}{8\pi f_a} 
     \bar{\tilde a} \gamma_5 \sigma^{\mu\nu}\tilde g^a
\, G^{a}_{\mu\nu},
\nonumber \\ 
 {\cal L}_{\tilde a\bino B}
       &=\frac{\alpha}{8\pi\cos^2\theta_W} \frac{C_{aY}}{f_a} 
     \bar{\tilde a} \gamma_5 \sigma_{\mu\nu}\bino 
           \,  
\left[
\cos\theta_W F^{\mu\nu}-\sin\theta_WZ^{\mu\nu}
\right],
\end{align}
where $C_{aY}=5/3$ if the number of $U$ quarks  
is the same as that of $D$ quarks~\cite{Hisano:1996ja,Freitas:2011fx},
and
 $\theta_W$ is the electroweak mixing angle.
%The axion-photon coupling depends on $z$ value, 
%$C_{a\gamma}=C_{aY}-2(4+z)/[3(1+z)]$.

At the LHC, axinos are produced as decay products of gluinos and Binos.
However
the decay rate $\Gm(\tilde{g} \to g \ax)$,
which is proportional to $m_{\tilde{g}}^3/f_a^2$,
is extremely suppressed by large value of $f_a \gsim 10^4 \gev$.
Other decay channels of the gluino
through strong and/or electroweak interactions
have much larger decay rates,
and thus make $\br(\tilde{g} \to \ax g)$
negligible.

The Bino NLSP is different.
Only allowed is
its decay into an axino:
no matter how small its decay rate is,
the branching ratio of $\br( \bino \to \ax \gm)$
is  almost 100\%.
The decay width of $\bino \to \tilde a \gamma$ is given by
\begin{align}
 \Gamma(\bino \to \tilde a \gamma)
              =\frac{\alpha^2}{128\pi^3}\frac{C_{aY}^2}{f_a^2\cos^2\theta_W}
                   m_{\bino}^3,
\end{align}
which leads to
the lifetime of Bino as
 \bea
\label{eq:tau}
  \tau_{\bino}\simeq 
          0.038\left(\frac{100 \gev}{m_{\bino}}\right)^3
          \left(\frac{f_a/C_{aY}}{10^6\gev}\right)^2\rm ns.
 \eea
%\begin{align}
% c\tau_{\bino}\simeq 1.2 \left(\frac{100 \rm GeV}{M_{\tilde B}}\right)^3\left(\frac{f_a/C_{aY}}{10^6 \rm GeV}\right)^2\rm cm.
% \end{align}
%For light speed, 1 ns is about 1/3 meter.
In the axino-LSP and Bino-NLSP scenario, therefore,
SUSY particle production
leads to the final states of diphoton and missing $E_T$, accompanied by SM particles. 
This process is phenomenologically identical to that of the GMSB model.

\begin{figure}
\begin{center}
\includegraphics[width=14cm]{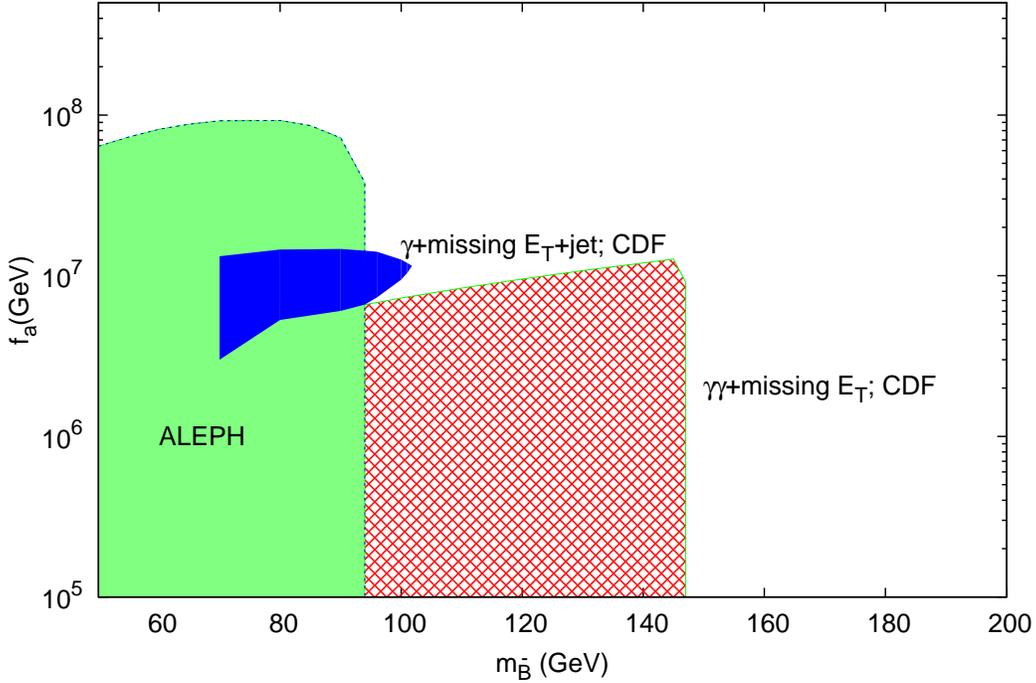}
\end{center}
\caption{
Excluded region in the parameter space of $(m_\bino,~f_a)$ 
by ALEPH $\gm\gm+\met$, CDF $\gm\gm+\met$ and CDF $\gm+\met+$jet data. 
We set $C_{aY}={5}/{3}$
and the benchmark scenario in Eq.~(\ref{eq:benchmark}). 
}
\label{fig:lep2}
\end{figure}

\begin{figure}
\begin{center}
\includegraphics[width=14cm]{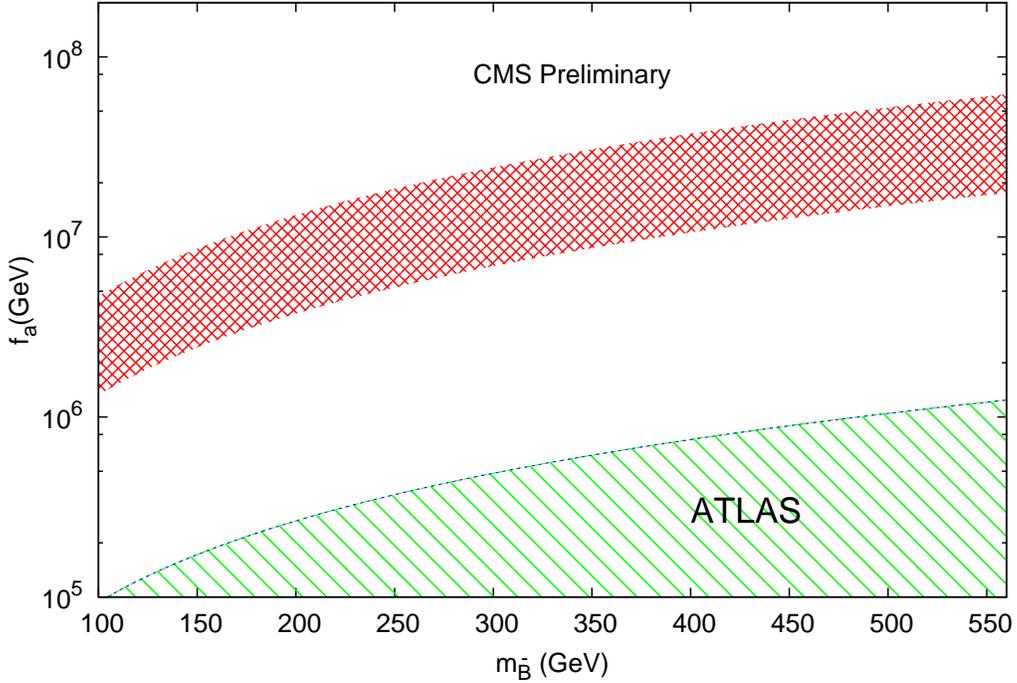}
\end{center}
\caption{
Excluded region in the parameter space of $(m_\bino,~f_a)$ 
by the ATLAS $\gm\gm+\met$ data and 
the CMS preliminary $\gm+\met+$jet data. 
We assume that the inclusive production cross section of Bino
is large enough for both ATLAS and CMS.
}
\label{fig:lhc}
\end{figure}

As comprehensively summarized in Fig.~\ref{fig:lep2} and \ref{fig:lhc},
we study the implications 
of the diphoton events plus missing energy
by the ALEPH, CDF, ATLAS and CMS experiments.
All of the results are consistent with the SM backgrounds.
There are two interpretations for this null result.
First is that the new physics scale is too high 
to yield an excess over the backgrounds.
The new physics cross section is too small.
An alternative interpretation is possible in our scenario.
Binos are produced enough
but they decay outside the detector so that
we do not see $\gm\gm+\met$ signal.
This interpretation is possible because of very high scale of $f_a$.
In this paper, we take the second interpretation.

We specify our model in more detail.
%We also set the mass spectrum of SUSY particles.
Since main production channels of the Bino pair 
at a hadron collider
are the cascade decays from gauginos,  gluinos, and squarks,
the results are sensitive to their mass parameters.
For simplicity, we take the following SUSY particle mass spectra:
\bea
\label{eq:benchmark}
m_\neut &\simeq& m_\cha \simeq 1.8 \,m_\bino,
\quad
m_{\tilde{\ell}_R} \simeq 1.1 \,m_\bino,
\quad
m_{\tilde{\ell}_L} \simeq 2.5 \,m_\bino,  \\ \no
m_\gno &\simeq& 800\gev,
\quad
m_{\tilde{q}} \simeq 1.5\tev.
\eea
The second lightest neutralino $\neut$ and the lightest charging $\cha$
are assumed to be Winos.
The gaugino and slepton mass relations with the Bino mass
are motivated from the GMSB SPS8 slope.
The gluino mass is from the recent CMS searches for the MSSM signal~\cite{gluinomass}.
Note that the slepton masses do not affect the bounds from the 
CDF, ATLAS, and CMS experiments.
One possible concern is that our benchmark scenario may have too light right-handed selectron. 
This could be excluded by the OPAL data on di-lepton plus missing energy, which provided, 
at 95\% C.L., the exclusion region of the right-handed selectron mass and the LSP mass up to 
$m_\bino = 80 \gev$~\cite{opal}. 
Our condition of $m_{\tilde{\ell}_R} \simeq 1.1 \,m_\bino$
is marginally allowed.

%ALEPH data
First, the ALEPH data put a very strong bound on $f_a$ 
for light Bino case.
The ALEPH group searched for the GMSB 
reaction $e^+e^-\to \bino\bino\to
\grno\grno\gamma\gamma$
at $\sqrt{s}=189- 209$ GeV
with a total integrated luminosity of $628 \pbi$~\cite{Heister:2002ut,Abdallah:2003np}. 
Crucial event selection criteria is to demand
photons not originating from the primary vertex of the interaction,
which is useful for longer Bino lifetime.
Our benchmark scenario in Eq.~(\ref{eq:benchmark})
guarantees large enough Bino direct production,
which is sensitive to the slepton mass.
More general slepton masses, but still not so heavy compared to the Bino mass,
result in similar results.
No excess of $\gm\gm+\met$ signal  over the SM backgrounds
can be allowed if the lifetime of the Bino is long enough.
If the Bino decays within the detection reach,
our scenario is not consistent with the null result,
which excludes small $m_\bino$ and small $f_a$.
In Fig.~\ref{fig:lep2}, we show the 
exclusion region 
denoted by ``ALEPH''.
The limited c.m. energy of the LEP2
covers the Bino mass only up to about 94 GeV.
In this small Bino mass range, however,
the bound on $f_a$ is very stringent:
$f_a < 10^8 \gev$ is mostly excluded.
The narrow hadronic axion window is completely closed 
for the light Bino case.

The CDF collaboration also presented
their analysis of $\gm\gm+\met+X$ data,
based on $2.6 \fbi$ total luminosity 
at $\sqrt{s}=1.96\tev$~\cite{Aaltonen:2009tp}. 
Here $X$ denotes other high-$p_T$ SM particles.
This search is motivated by
the GMSB model.
Significant improvement on the search sensitivity
was made by a new photon timing system 
in the EM calorimeter, which measures the arrival time of photons \cite{Toback}.
With the tracking device measuring the position and time
of the $\ppb$ primary collision,
this timing system probes the Bino life time $\tau_\bino$ up to 2 ns.
%For $\tau_\bino \gsim 2\ns$, ambiguity 
%from the next collision arises .

Longer Bino lifetime region 
in $0< \tau_{\bino} <40\,\rm ns$
is also covered by CDF experiment through $\gm+{\rm jet}+\met$~\cite{Aaltonen:2008dm}.
This final state is motivated by their reference model,
the GMSB SPS8 point.
Here the main Bino production channels through gaugino decays
are associated with prompt taus whose decays can be identified 
as jets.
Photon arrival time is measured by a timing system in the ECAL.
Final states are a high-$E_T$, isolated,  
and {\em delayed} photon with large $\met$ and  a high-$E_T$ jet.
Null results with the integrated luminosity of $570 \pbi$ have been reported.

In our benchmark scenario,
gluinos and squarks are  
too heavy to 
have sizable production cross sections at the Tevatron.
The produced Binos are mainly from the cascade decays of the gaugino pair production
of $\neut \cha$ and $\chap\cham$.
The inclusive Bino pair production is large enough for $m_\bino \lsim 150 \gev$.
If the Bino decays outside the reach of
the photon timing system,
our scenario is allowed.
Heavier Bino case, corresponding to heavier neutral and charged Winos,
leads to too small cross section:
the CDF data cannot constrain the model.
Based on two kinds of CDF data,
we exclude the parameter space
of $(m_{\bino},f_a)$ in Fig.~\ref{fig:lep2}.
The CDF data put also strong bound
on $f_a \lsim 10^7\gev$ 
for $100\gev \lsim m_\bino < 150\gev$.

%LHC
The CMS and ATLAS 
collaborations~\cite{Aad:2011kz, arXiv:1111.4116,arXiv:1103.0953}
have also reported 
their search for diphoton and missing transverse energy.
With the integrated luminosity of 1 fb$^{-1}$ at $\sqrt{s}=7$ TeV,
the ATLAS group presented their analysis 
of $\gm\gm+\met$ final states~\cite{arXiv:1111.4116}.
The ATLAS search is limited for $c\tau_\bino < 0.1\;$mm.
At the LHC, 
multiple collisions from high luminosity  
make it very challenging to measure
the timing separation between the primary collision and the photon arrival.
In our model with $\sqrt{s}=7\tev$,
the gluino pair production becomes also important especially when the Bino mass
is large.
Following our basic interpretation,
the decay of a Bino outside the reach of $c\tau_\bino < 0.1\;$mm is allowed.
This excludes the $(m_\bino,f_a)$ parameter space as in Fig.~\ref{fig:lhc}.

The CMS collaboration also presented their
preliminary data 
of $\gm+\met+{\rm jet}$ event at integrated luminosity of 2.1 fb$^{-1}$~\cite{EXO-11-067}. 
The CMS group applied the
photon conversion impact parameter method 
and searched for long-lived  ($2\, {\rm cm}<c\tau_{\bino}<25\,$cm) 
Bino decaying into a photon and a gravitino. 
The {\em transverse} impact parameter of a photon to the beam line is measured,
which is a robust observable when the true primary vertex is not known
because of multiple collisions.
As in the CDFII delayed photon detection, this method also measures the delayed photon 
from long-lived Bino decay.
The CMS preliminary results set the upper limits on the Bino inclusive 
production cross section $\sigma<(0.12-0.24)\,\rm pb$. 
In our benchmark model, 
gluino pair production is the main channel with the total cross section about $\sim 100 \fb$.
Based on these upper limits,
we exclude the parameter space of
$m_\bino$ and $f_a$,
as in Fig.~\ref{fig:lhc}.
The band structure of the exclusion region is from the limitation of the CMS photon
conversion impact parameter method.
The ATLAS and CMS data extend the search for larger Bino mass.
The bound on $f_a$ is quite significant,
which closes a large portion of hadronic axion window
 $f_a \sim 10^6\gev$ especially when the Bino is heavy.

In conclusion, we obtain the improved bound on the axion
decay constant
and the Bino mass in the axino-LSP and Bino-NLSP model 
from the ALEPH, CDF II and early LHC data 
on $\gm\gm+\met$ signal.
We take a benchmark scenario specifying the SUSY particle masses.
Light Bino case with mass below about $150\gev$
has a strong constraint from the ALEPH and CDF data:
the region of $f_a<10^7\gev$ is 
completely excluded.
The hadronic axion window 
is almost closed in this region.
The $1 \fbi$ ATLAS data exclud a large portion of
$f_a < 10^6$ GeV region up to $m_{\bino}\simeq 700\gev$.
 And $2.1 \fbi$ CMS data exclude a region around $ f_a \sim 10^7$ GeV. 
This bound is much stronger 
than the previous laboratory bound ($f_a>10^4\gev$).

\acknowledgments
The work  of SC and KYL is supported by WCU program through the KOSEF funded
by the MEST (R31-2008-000-10057-0).
The work of KYL is also supported by
the Basic Science Research Program 
through the National Research Foundation of Korea (NRF) 
funded by the Korean Ministry of
Education, Science and Technology (2010-0010916).
The work of JS is supported by
the National Research Foundation of Korea (NRF) 
funded by the Korean Ministry of
Education, Science and Technology (2011-0029758).

%%%%%%%%%%%%%%%%%% References
%%%%%%%%%%%%%%%%%%%%%%%%%%%%%%%%%%%%%%%%%%%%%%%%%%%%%%%
\def\PRD #1 #2 #3 {Phys. Rev. D {\bf#1},\ #2 (#3)}
\def\PRL #1 #2 #3 {Phys. Rev. Lett. {\bf#1},\ #2 (#3)}
\def\PLB #1 #2 #3 {Phys. Lett. B {\bf#1},\ #2 (#3)}
\def\NPB #1 #2 #3 {Nucl. Phys. B {\bf #1},\ #2 (#3)}
\def\ZPC #1 #2 #3 {Z. Phys. C {\bf#1},\ #2 (#3)}
\def\EPJ #1 #2 #3 {Euro. Phys. J. C {\bf#1},\ #2 (#3)}
\def\JHEP #1 #2 #3 {JHEP {\bf#1},\ #2 (#3)}
\def\IJMP #1 #2 #3 {Int. J. Mod. Phys. A {\bf#1},\ #2 (#3)}
\def\MPL #1 #2 #3 {Mod. Phys. Lett. A {\bf#1},\ #2 (#3)}
\def\PTP #1 #2 #3 {Prog. Theor. Phys. {\bf#1},\ #2 (#3)}
\def\PR #1 #2 #3 {Phys. Rep. {\bf#1},\ #2 (#3)}
\def\RMP #1 #2 #3 {Rev. Mod. Phys. {\bf#1},\ #2 (#3)}
\def\PRold #1 #2 #3 {Phys. Rev. {\bf#1},\ #2 (#3)}
\def\IBID #1 #2 #3 {{\it ibid.} {\bf#1},\ #2 (#3)}

\end{document}